# Investigating Blended Math-Science Sensemaking with a diverse undergraduate population

Leonora Kaldaras, Carl Wieman

Stanford Graduate School of Education

**Abstract**

**Background**: Blended mathematical sensemaking in science ("MSS") involves deep conceptual understanding of quantitative relationships describing scientific phenomena. Previously we developed the cognitive framework describing proficiency in MSS across STEM disciplines. The framework was validated with undergraduate students from dominant backgrounds (White, middle class) using assessment built around PhET simulations. In this study we investigate whether the framework can characterize engagement in MSS among undergraduate students from diverse backgrounds and identify potential differences between the two populations. This study provides insights on how to better support diverse students in building MSS skills.

**Results**: the framework is effective in characterizing engagement in MSS by undergraduate students from diverse backgrounds and largely functions as a learning progression. We have also uncovered a distinct pattern of engagement in MSS reflected in students successfully developing the mathematical relationship describing their observations without engaging in a type of MSS focused on quantitative pattern identification. Unlike students from dominant backgrounds, diverse students leverage lower level MSS to develop the formula for the phenomenon in question. Further, diverse students use PhET simulations to make sense of the phenomenon more and are more successful in using the simulations to find the correct mathematical relationship compared to students from dominant backgrounds. Finally, Math preparation has a stronger effect at level 1 compared to levels 2 and 3 of the framework.

**Conclusion:** the framework can guide the development of instructional and assessment strategies to support students from diverse backgrounds in building MSS skills. The framework helped identify specific types of MSS that need to be supported to facilitate the transition to the highest levels of the framework among these students, and these types of MSS turned out to be the same for both dominant and diverse groups of students. Further, PhET simulations provide a suitable and effective learning environment for supporting engagement in and learning of MSS skills, and their capabilities should be leveraged for designing learning experiences in the future. Finally, level 1 students should be supported in developing a deeper understanding of Math as well as integration of Math and Science when making sense of phenomena.

**Introduction**

The ability to explain scientific phenomena mathematically by integrating scientific and mathematical reasoning is reflective of deep science understanding (Redish, 2017; Zhao & Schuchardt, 2021; Kuo et al., 2013). This foundational cognitive process is called *blended math-science sensemaking* (MSS) and lies at the heart of scientific thinking (Redisch, 2017; Zhao & Schuchardt, 2021). MSS is characterized by the ability to integrate the relevant math and science ideas to make sense of natural phenomena mathematically. It is through this blending of math and science that we can develop highly precise and predictive models of the natural world. While the value of blended MSS has been well recognized, traditional STEM courses don't effectively support the development of this skill (Becker & Towns, 2012; Bing & Redish, 2007; Taasoobshirazi & Glynn, 2009; Tuminaro & Redish, 2007).

The majority of students don't naturally engage in blended MSS, and therefore need carefully scaffolded instructional support for developing this ability (Kaldaras & Wieman, 2023a). Since MSS involves the ability to integrate the relevant math and science resources, developing effective learning environments for supporting MSS calls for a closer examination of student struggles in using these resources together to explain phenomena. Moreover, students from various academic and cultural backgrounds may exhibit diverse patterns of integration of math and science resources, which are important to consider when developing learning activities aimed at supporting all students. For example, prior research has shown that insufficient math preparation among students from backgrounds historically underrepresented in science is one of the barriers to their building deep understanding of science topics that require understanding of relevant science and math ideas, such as stoichiometry, which requires understanding of the concept of mole and proportional reasoning (Ralph & Lewis, 2018). However, with sufficient instruction on the topic, students from underrepresented backgrounds do equally as well or outperform their peers from dominant backgrounds (Ralph & Lewis, 2018). Further, while math preparation is important for succeeding in STEM, prior research has also found that it is not sufficient for effective engagement in blended MSS (Kaldaras & Wieman, 2023a). Rather, it is important to support students in developing the ability to *integrate* math and science dimensions to effectively engage in MSS (Kaldaras & Wieman, 2023a). Therefore, incorporating supportive activities that target MSS in science courses can potentially benefit all students and could remove barriers for underrepresented students.

Designing effective instructional systems for learning in any domain requires understanding of how proficiency in that domain develops (National Research Council [NRC], 2000, 2012 a, b). Proficiency refers to describing what mastery looks like in a given domain. Cognition models such as learning progressions (LPs) allow for aligning curriculum, instruction, and assessment with the purpose of helping students achieve higher proficiency in a given concept (Duschl et al., 2007). LPs represent a continuum of increasingly sophisticated ways of thinking about a given concept that develops across a broad, defined period (Duschl et al., 2007). LPs are usually grounded in research on how students learn ideas associated with scientific constructs of interest and specific logic of a given discipline.

Since LPs reflect a detailed qualitative description of student understanding at various levels of sophistication, they represent a useful tool for helping design more equitable learning systems by helping adjust the learning supports to the needs of individual students (Delgado & Morton, 2012). Therefore, using LPs to guide the development of learning systems can potentially benefit students with different levels of prior preparation by helping adjust instruction to their individual cognitive level (Delgado & Morton, 2012; Kaldaras & Krajcik, in press). However, this is only possible if the LP design reflects the diversity of student thinking at all levels of sophistication (Delgado & Morton, 2012; Kaldaras & Krajcik, in press).

Complexity of student responses does not always map well into specific LP levels. If "atypical" response patterns are ignored, this also perpetuates educational inequalities by prioritizing the researchers' and teachers' views on canonical scientific knowledge (Kang & Furtak, 2021). Therefore, designing and validating LPs should involve careful consideration of various ways students from different backgrounds think about a given topic at all levels of sophistication. Historically, however, LP validation studies have been conducted mostly with students from dominant White rural or suburban middle class backgrounds (Delgado & Morton. 2012). Moreover, the majority of LP studies did not focus on characterizing and incorporating all student ideas into the LP, but rather focused on ideas shared by most students (Delgado & Morton. 2012). This limitation of the LP studies poses significant challenges for using such LPs to guide instruction and support students from diverse backgrounds.

MSS has been studied in various fields of science (Bing & Redish, 2007; Tuminaro & Redish, 2007; Ralph & Lewis, 2018; Lythcott, 1990; Schuchardt & Schunn, 2016; Hunter et al., 2021; Zhao & Schuchardt, 2021). While it is an important component of expert-like understanding and expert mental models, there has been limited work on defining and validating a cognitive model describing proficiency in this construct across STEM fields. Initial work was done by Zhao and Schuchardt (2021) that proposed a theoretical cognitive model for independent mathematics and science dimensions outlining increasingly sophisticated proficiency levels. Their model is grounded in a review of relevant literature across different fields (including math, physics, chemistry, biology) and represents mathematical and scientific sensemaking as separate dimensions. Building off this work, we developed and validated a cognitive framework for MSS proficiency levels (Kaldaras & Wieman, 2023a). We have found that MSS proficiency is largely independent of the specific disciplinary context: most students engaged in MSS consistently in the same level across three disciplinary scenarios including Physics, Chemistry and an interdisciplinary scenario (Kaldaras & Wieman, 2023a). In addition, the level assignments seemed to be reasonably well related with SAT Math scores below 650 but did not distinguish well for scores above 650 suggesting a strong integration of Math and Science dimensions in the context of MSS (Kaldaras & Wieman, 2023a). We also found that the cognitive framework largely functions as a LP as reflected in student response patterns. Specifically, students tended to engage in MSS at all the levels of sophistication described by the framework up until a particular highest attained level. They typically were able to engage in MSS at all the levels below their highest demonstrated one (Kaldaras & Wieman, 2023a). This property of our validated cognitive framework for MSS suggests that it can potentially be used to guide and adjust instruction to the needs of individual learners. We have already shown that the learning sequences for introductory Chemistry and Physics courses grounded in the cognitive framework help support students in transitioning from the lowest to the highest levels of the framework (Kaldaras & Wieman, 2023b). However, prior work was done based on interviews and instructional activities with students recruited from large Western US universities (one private and one state) with predominantly White middle class student populations. Therefore, our student sample in the previous studies did not incorporate students from diverse academic and cultural backgrounds and as a result may not reflect the ways these more diverse students engage in MSS. This is the limitation we are addressing in this study.

This work investigates the ways in which students from very different institutions, predominantly 2 year colleges with large enrollments of students from underrepresented backgrounds engage in MSS . We recruited freshmen students from community colleges with more than 50% underrepresented minority students and University of Puerto Rico, which is a historically Hispanic-serving institution. This study focuses on answering the following research questions (RQs):

*RQ1: Can the previously validated cognitive framework for MSS be used to characterize the different ways undergraduate students from various institutions and underrepresented backgrounds engage in MSS?*

*RQ 2: Do undergraduate students from various institutions and underrepresented backgrounds demonstrate similar patterns of engagement in MSS as the university students from dominant backgrounds?*

We address these RQs by conducting interviews with the targeted student population using previously described approach and protocol (Kaldaras & Wieman, 2023a).

An important aspect of the approach we developed in the previous study was designing the interview protocol for probing the levels of the MSS framework in the context of relevant PhET simulations. Specifically, one of the key features of the sensemaking process is its dynamic nature focused on continuously revising an explanation based on new evidence to figure something out (Odden, Russ, 2019). The dynamic nature of PhET simulations provides a unique and suitable environment for assessing MSS skills. They offer a dynamic and interactive assessment environment that allows for accumulation of new evidence and feedback associated with changing parameters of the system in question. This supports revisions of explanations by calling on blended understanding of the scientific concepts and the underlying mathematical relationships. In the context of MSS, the relevant mathematical equations represent processes described by specific variables. Simulations, in turn, represent a physical behavior with certain variables that control that behavior. The simulation allows learners to explore how the behavior depends on different variables, both qualitatively and quantitatively, therefore providing a meaningful context for engaging in MSS. Simulations provide a simplified (but not too simplified) system for exploring the mathematical complexity of the phenomenon described in the simulation. These features of simulations were the reason for choosing PhET simulations as the assessment context for testing our theoretical cognitive model for MSS.

  The results of the prior study with students from dominant backgrounds showed that PhET simulations indeed permitted meaningful engagement in MSS at all the levels described by the cognitive framework (Kaldaras & Wieman, 2023a). Moreover, a different prior study on designing instructional sequences grounded in the cognitive framework and using relevant PhET simulations showed that the simulations that allow students to explore specific quantitative patterns among the relevant variables are more effective in helping students transition to higher levels of the framework than those that only allow for qualitative exploration (Kaldaras & Wieman, 2023b). However, both qualitative and quantitative PhET simulations seemed to be effective in supporting engagement in MSS and student progress to the higher framework levels (Kaldaras & Wieman, 2023b). Therefore, prior studies indicate that PhET simulations offer a promising tool in supporting student engagement and proficiency development in an otherwise complex process of MSS. An alternative- traditional static instruction that leverages text or drawings or the use of real equipment to support learning such as blended MSS, while possible in principle, imposes an additional cognitive load on learners (Madden et al., 2020; Holmes, 2020; Wieman & Perkins, 2005). PhET simulations, on the other hand, are designed to help avoid excessive cognitive load (Adams et al., 2008 a, b) and can support tasks that are sufficiently scaffolded and simple enough to allow for systematic engagement in MSS (Kaldaras & Wieman, 2023b). In this study we aim to investigate whether PhET simulations offer similar degree of support for engagement in MSS to students from underrepresented backgrounds as they do to students from dominant backgrounds and whether they are effective in supporting the opportunity to learn blended MSS skills among these students. We therefore will be answering the following RQ:

*RQ 3: Do PhET simulations help undergraduate students from different institutions and underrepresented backgrounds meaningfully engage in and learn MSS across STEM disciplines?*

We will use the results of interviews conducted with individual students around relevant PhET simulations to answer this RQ. We will also discuss the implications of our findings for broad application of the cognitive framework for MSS and use of PhET simulations to support this process.

**Theoretical Framework**

*LPs as tools to create more equitable learning environments*

LPs describe a path students can follow when developing a deeper understanding of a given topic (Duschl et al., 2007). LPs are grounded in cognitive constructivist perspectives on learning emphasizing that individuals construct their understanding through interactions with physical and social environment by building on their prior knowledge (Duschl et al., 2007; NRC, 2000). LPs are moderated by curriculum and instruction- that is, LPs are only effective in guiding and supporting student learning if students have opportunities to learn the ideas reflected in the LP (Duschl et al., 2007; Kaldaras et al., 2021; Stevens et al., 2009). In other words, whether the learner shows progress along the LP over time is closely related to the affordances of the learning system. This view on learning with LPs is compatible with the sociocultural perspective on equity and opportunity to learn, which suggests that equitable learning environments should afford individualized attention to learners in order to effectively compensate for past lack of learning opportunities and therefore promote social justice (Delgado & Morton, 2012). Moreover, to be effective these affordances should allow learners to perceive them as feasible (Gee, 2008; Kang, 2018). Therefore, LPs have the potential to help guide the development of equitable learning systems if paired with learning environments that support equitable opportunities to learn for a wide range of students.

In order for the LP to be useful for diverse learners, it is critical that the LP reflects a wide range of student thinking. As mentioned above, a significant limitation of modern LP research is that most LPs reflect the most common ways of thinking about a given topic, and don't offer a meaningful way to deal with atypical response patterns that don't align well with a given LP (Kang & Furtak, 2021; Delgado & Morton, 2012). This limitation of the LP design has been recognized to perpetuate deficit perspective by viewing atypical student responses as flawed and prioritizing researchers' and teachers' views on scientific knowledge rather than centering students' experiences (Kang & Furtak, 2021).

This study aims to address this limitation of LP research. First, this study is grounded in a previously validated cognitive model for MSS that was specifically focused on incorporating a wide range of student thinking grounded in students' experiences and observations. Second, in our development we considered lower-level student responses as steppingstones to engaging in more sophisticated types of MSS given the necessary resources (Kaldaras & Wieman, 2023a). Third, in the current study we account for a wider range of student engagement in MSS by targeting students from a variety of academic and cultural backgrounds.

*Prior Work: Validated Cognitive Framework for MSS*

The levels of an LP are usually expressed as learning performances that summarize what students should be able to do with the scientific knowledge they have (Reiser et al., 2003). In the previous study we have developed the cognitive framework which appears to function as a LP for MSS. The LP is shown in Table 1. The LP describes increasingly sophisticated ways of engaging in MSS as students are working towards developing a mathematical formula describing the scientific phenomenon or building a deeper understanding of the known formula. The model consists of three broad levels: qualitative (level 1), quantitative (level 2) and conceptual (level 3). Each broad level consists of three sub-levels: "Description", "Pattern", and "Mechanism". At the qualitative level students can't develop the exact mathematical relationship describing the scientific phenomenon in question, but they can identify the relevant variables ("Description"), qualitative patterns among the variables ("Pattern") and describe a qualitative causal mechanism of the phenomenon ("Mechanism"). At the quantitative level students can identify numerical values of the relevant variables ("Description"), quantitative patterns among the variables ("Pattern") and develop a mathematical relationship describing the phenomenon and justify the equation using numerical values of the variables ("Mechanism"). At the conceptual level students justify the scientific need to include all unobservable variables and constants into the mathematical relationship ("Description"), justify the mathematical relationship by relating the observed quantitative patterns to specific mathematical operations ("Pattern") and describe the causal mechanism of the phenomenon reflected in the equation structure ("Mechanism").

Table 1. *Theoretical Blended Math-Sci Sensemaking Framework.*

| | | |
|---|---|---|
| **1 Qualitative** | Description | Students can use observations to identify which measurable quantities (variables) contribute to the phenomenon.<br>*Example: force and mass make a difference in the speed of a car.* |
| | Pattern | Students recognize patterns among the variables identified using observations and can explain *qualitatively* how the change in one variable affects other variables, and how these changes relate to the scientific phenomenon in question.<br>*Example: the smaller car speeds up more than the big car when the same force is exerted on both.* |
| | Mechanism | Students demonstrate *qualitative* understanding of the underlying causal scientific mechanism (cause-effect relationships) behind the phenomenon based on the observations but can't define the exact mathematical relationship.<br><u>*Example*</u>: *it is easier to move lighter objects than heavy objects, so exerting the same force on a lighter car as on a heavy car will cause the lighter car to speed up faster.* |
| **2 Quantitative** | Description | Students recognize that the variables identified using the observations provide measures of scientific characteristics and can explain *quantitatively* how the change in one variable affects other variables (but not recognizing the quantitative patterns yet), and how this change relates to the phenomenon in question. Students not yet able to express the phenomenon as an equation.<br><u>*Example*</u>: *recognizing that as variable A changes by 1-unit, variable B changes by 2 units.* |
| | Pattern | Students *recognize quantitative patterns* among variables and explain *quantitatively (in terms of an equation or formula)* how the change in one parameter affects other parameters, and how these changes relate to the phenomenon in question. Students not yet able to relate the observed patterns to the operations in a mathematical equation and can't develop the exact mathematical relationship yet.<br><u>*Example*</u>: *recognizing linear and inverse relationships* |
| | Mechanism | Students can explain *quantitatively* (express relationship as an equation) for how the change in one |

| | | | |
|---|---|---|---|
| | | | variable affects other variables based on the quantitative patterns derived from observations. Students include the relevant variables that are not obvious or directly observable. Students not yet able to explain conceptually why each variable should be in the equation beyond noting that the specific numerical values of variables and observed quantities match with this equation. Students cannot explain how the mathematical operations used in the equation relate to the phenomenon, and why a certain mathematical operation was used. Students can provide causal account for the phenomenon.<br>*Example*: In $F_{net}=m*a$, multiplication makes sense because when applied force on the mass of 50 kg increases from 10 to 20 N, acceleration increases by 2. That only makes sense for a multiplication operation. |
| 3<br>Conceptual | | Description | Students can describe the observed phenomenon in terms of an equation, and they can explain why all variables or constants (including unobservable or not directly obvious ones) should be included in the equation. Students not yet able to explain how the mathematical operations used in the formula relate to the phenomenon.<br>*Example*: In $F=m*a$, the F is always less than applied force by specific number, so there must be another variable subtracted from $F_{applied}$ to make the equation work. The variable involves the properties of the surface. So, the equation should be modified: $F_{applied}-(variable)=m*a$ |
| | | Pattern | Students can describe the observed phenomenon in terms of an equation, and they can explain why all variables or constants (including unobservable or not directly obvious ones) should be included in the equation. Students not yet able to provide a causal explanation of the equation structure.<br>*Example*: In $F_{net}=m*a$, multiplication makes sense because as applied force on the same mass increases, acceleration increases linearly, which suggests multiplication. |
| | | Mechanism | Students can describe the observed phenomenon in terms of an equation, and they can explain why all variables or constants (including unobservable or not directly obvious ones) should be included in the equation. Students can fully explain how the mathematical operations used in the equation relate to the phenomenon in questions and therefore demonstrate *quantitative* conceptual understanding. Here, the conceptual understanding is how mathematical relationships represent numerical dependencies.<br>*Example*: since greater acceleration is caused by applying a larger net force to a given mass, this shows a positive linear relationship between a and $F_{net}$, which implies multiplication between m and a in the equation, or $F_{net}=m*a$. |

Notice that the sub-levels of the framework reflect a wide range of ways students can engage in MSS from the least to the most sophisticated. Moreover, the sub-level description is grounded in the MSS process that students engage in based on their observations of the relevant phenomena (or, by interacting with the simulation describing the phenomenon). For example, sample responses in Table 1 reflect the type of MSS students engage in when describing how the force applied to objects of various masses affects their acceleration (Newton's Second Law). Students can provide a very simple, qualitative account of relationships among the different variables as reflected in sample responses for level 1 ("qualitative"). At the next levels, they start engaging with numbers and provide a quantitative account of the phenomenon at varying levels of sophistication as reflected in sample responses for level 2 ("quantitative"). Notice that the types of blended MSS described for level 2 build meaningfully on blended MSS demonstrated at level 1. In other words, the framework offers a path to leveraging the level 1 type of MSS to help students attain level 2. For example, once students have identified qualitative patterns using their observations such as that smaller cars will accelerate more with the same applied force compared to heavier car (level 2 (Qualitative) "Pattern" example), instructional supports can be provided to students to investigate by exactly how much more a lighter car of mass X will accelerate compared to a car of mass B, which will allow learners to engage in level 2 (Quantitative) "Pattern" type of MSS. Depending on how much experience students have in noticing the quantitative patterns among the data points, they might need multiple data sources and various scaffolds to help them build connections between the qualitative observations and quantitative patterns in the data. The main instructional goal would be to help them identify the inverse linear relationship between mass and acceleration and the direct linear relationship between applied force and acceleration. Similarly, all other sub-levels are developed to build on each other and support student engagement in MSS from the most basic to the most sophisticated one. Each sub-level is grounded in student ideas that can be generated from observations rather than disciplinary or scientifically normative and accurate ideas. Therefore, our LP for MSS centers the view on proficiency in MSS in student experiences and observations, avoiding the deficit perspective on MSS and offering a path to adjusting instruction meaningfully to the needs of individual learners starting with their prior knowledge.

Moreover, the framework views all types of MSS (including those consistent with lower levels) as productive and treats them as stepping stones to engaging in higher level MSS, which reflects an anti-deficit perspective on sensemaking (Adiredja, 2019). In the current study we aim to further test the utility of this framework in helping characterize the different types of MSS that students from various underrepresented backgrounds demonstrate and compare the results to the previous study conducted with students from mostly dominant backgrounds. This further evaluates the effectiveness of this framework for guiding the design of more equitable individualized learning environments that support students from a wide range of backgrounds in developing MSS skills.

*Computer Simulations as tools to support opportunity to learn for MSS*

As mentioned above, LPs are only effective in helping students transition to higher levels if students have opportunities to learn the ideas outlined in the LP. Creating feasible opportunities to learn for a diverse group of learners is a challenging task because it is not enough to simply expose learners to certain information (Gee, 2008). Rather, it is important to focus on the relationship between the learner and the learning environment and consider whether the learning environment affords the types of learning opportunities that a given group of individual learners can perceive as feasible to help them engage in the learning process (Gee, 2008). As discussed above, MSS is a complex cognitive process that involves continuous accumulation of evidence associated with changing the parameters of the system and revising an explanation based on new evidence to figure something out (Odden & Russ, 2019). For the reasons given, this makes computer simulations such as PhET simulations a promising alternative for engaging students in MSS. By limiting cognitive load and providing intuitive visual representations, they are effective for students with a wide range of educational backgrounds. They provide students with the freedom to manipulate various parameters, but also focusing their exploration only on the aspects that are the most relevant for understanding how relevant variables affect the observations of the phenomena (Adams et al., 2008 a, b). They represent a promising way of tackling the issue of the power of representation, which states that some forms of representation are more effective for some learners than others depending on their prior preparation (Gee, 2008) by offering a simplified (but not too simplified) system for exploring all the relevant mathematical aspects of the system and relating them to the overarching phenomenon without the need for any prior knowledge. In short, PhET simulations offer a promising tool for helping students from a wide range of backgrounds authentically engage in blended MSS by lowering the cognitive load, minimizing the requirement of prior knowledge, and offering a widely accessible representation of all the resources needed for meaningful engagement in blended MSS.

**Methods**

*Data*

We used interviews with individual students as the source of data for the current study. The interview protocol and the overall study design was previously published (Kaldaras &Wieman, 2023a). Briefly, each participating student completed two 1-hour long interviews focusing on a Physics and Chemistry disciplinary scenario respectively. Both interviews were structured around the relevant PhET simulations and used the same previously published interview protocol (Kaldaras & Wieman, 2023a). The overarching task for both interviews was to develop a mathematical relationship describing observations in the simulation. For the Physics disciplinary scenario students were exploring the Acceleration[1] simulation and were intended to develop the equation for Newton's Second Law ($F_{applied}-F_{friction}=m \times a$). For the Chemistry scenario students were exploring the Beer's law[2] simulation and were intended to develop the equation for Beer's law (Absorbance=concentration × MAC × container width), where MAC is molar absorption coefficient. During the interview, students were given time to explore the simulation first, and then were asked a series of questions to probe all the levels of the cognitive framework shown in Table 1. The interview protocol is provided in the Appendix. Briefly, students were asked to identify the variables relevant for characterizing the phenomenon mathematically, qualitative patterns among the variables and propose a qualitative causal mechanism (probing level 1 (Qualitative) "Description", "Pattern" and "Mechanism" respectively). Further, students were asked to identify relevant numerical values for the variables, quantitative patterns among the variables and propose a mathematical relationship describing their observations (probing level 2 (Quantitative) "Description", "Pattern" and "Mechanism" respectively). Further, students were asked to explain why all the variables and constants should be included in the proposed formula (probing level 3 (Conceptual) "Description"). Finally, in order to differentiate among level 2 (Quantitative) "Mechanism" and level 3 (Conceptual) "Pattern" and "Mechanism" students were asked to justify their proposed formula and describe a causal mechanism reflected in their formula. We will show the types of student responses consistent with each of these sublevels in 5the results section.

---

[1] https://phet.colorado.edu/sims/html/forces-and-motion-basics/latest/forces-and-motion-basics_en.html?screens=2
[2] https://phet.colorado.edu/sims/html/beers-law-lab/latest/beers-law-lab_all.html?screens=2

     If students struggled to develop the mathematical relationship after interacting with the simulation, they were also provided with the explicit quantitative data relevant to the simulation (see Appendix for the data that was provided for each interview). If students were still struggling to formulate a mathematical relationship, they were provided with a multiple-choice list of possible formulas and asked to pick the most likely one and justify their choice (see Appendix).

*Student Sample*

     We recruited the students by sending out emails to professors teaching introductory STEM courses in community colleges and deans of various STEM colleges from the list provided by Bill and Melinda Gates (BMG) foundation. We specifically targeted institutions with high percentages (50% and above) of enrollment of students from underrepresented minority groups (% URM enrollment - 3yr avg column in BMG foundation data set provided to us). We contacted only non-profit community colleges (2-year institutions) from BMG foundation list. In the email we asked to share information about paid interview opportunities with the PhET simulations project in a variety of introductory STEM courses. We have contacted over 50 professors and administrators in over 50 institutions and received a few positive responses indicating that the information was shared with students at the corresponding institution. We have also reached out to professional acquaintances in the University of Puerto Rico. As a result, we recruited students from the institutions shown in Table 2.

*Table 2. Educational Institutions attended by student interview volunteers.*

| Institution | Location | Number of Students | % URM enrollment - 3yr avg |
|---|---|---|---|
| Long Beach City College | CA | 3 | 70 |
| Bakersfield College | CA | 3 | 71 |
| Santa Fe Community College | NM | 1 | 54 |
| University of Puerto Rico | PR | 5 | NA |

A total of 126 students from the institutions shown in Table 2 completed the volunteer form that asked them about their academic background and identification with specific minority groups. When picking students for interviews, we ensured that students who participated in interviews were freshmen enrolled in introductory Chemistry and/or Physics courses. This ensures that we target the incoming freshmen population segment of that diverse undergraduate population. We recruited a total of 12 students, all of which completed two interviews described above. Each student was compensated with a $20 Amazon gift card for each interview they completed. Group-level student demographics are shown in Table 3.

*Table 3. Group-level student Demographics*

| **Underrepresented group type** | **Number of self-identified students** |
|---|---|
| Individuals with learning differences | 1 |
| Individuals living on low income | 7 |
| LGBTQ+ | 4 |
| Women | 5 |
| LatinX | 7 |
| Person of Color | 4 |
| Indigenous | 1 |
| Black | 1 |
| Identify in two groups | 6 |
| Identify in three groups | 4 |
| Identify in four groups | 1 |

Individual-level demographics is shown in Table 4. Academic background including the highest

Biology, Physics, Chemistry and Math courses taken for each participant is shown in Table 5. If the participant had their SAT Math score available or if they remembered an approximate score, that information is shown under the corresponding column.

*Table 4. Demographics background for each participant.*

| Student | Low income | Person of color | Learning differences | LGBTQ+ | LatinX | Indigenous | Black | Women |
|---|---|---|---|---|---|---|---|---|
| 1 | Yes |  | Yes | Yes |  |  |  |  |
| 2 | Yes | Yes |  |  |  |  |  |  |
| 3 |  |  |  |  | Yes | Yes |  |  |
| 4 |  |  |  |  | Yes |  |  | Yes |
| 5 | Yes | Yes |  |  |  |  |  | Yes |
| 6 |  |  |  |  | Yes |  | Yes | Yes |
| 7 | Yes |  |  | Yes |  |  |  |  |
| 8 |  |  |  |  | Yes |  |  |  |
| 9 | Yes | Yes |  |  |  |  |  |  |
| 10 | Yes |  |  |  | Yes |  |  | Yes |
| 11 |  |  |  | Yes | Yes |  |  |  |
| 12 | Yes | Yes |  | Yes | Yes |  |  | Yes |

*Table 5. Academic Background for Each Participant*

| Student | Biology | Physics | Chemistry | Math | Math SAT | Major | English |
|---|---|---|---|---|---|---|---|
| 1 | Intro Bio | none | HS Chem | College Stats. | ~500 | Diagnostic medical imaging | Native |
| 2 | HS Bio | Phys. II* | HS Chem | Differential equation and linear algebra | none | Not identified | Native |
| 3 | Intro Bio | Phys. II* | Gen. Chem. I | Calculus III | none | Computer science | Native |
| 4 | Intro Bio | none | Gen Chem. II | Calculus I | ~400 | Interdisciplinary studies | ELL |
| 5 | Intro Bio | none | Pre. College | Calculus I/College | 490 | Nutrition | Native |

| | | | Chem. | Stats. | | | |
|---|---|---|---|---|---|---|---|
| 6 | Intro Bio | HS Phys. | Gen. Chem. I | Pre-Calc. | 500-600 | Chemistry | Native |
| 7 | Intro Bio | none | Gen. Chem. I | Calculus I/College Stats. | none | Occupational Therapy (MS) | Native |
| 8 | Intro Bio | Phys. II | HS Chem. | Higher than College Stats. | 680 | Electrical Engineer | Native |
| 9 | Intro Bio | HS Phys. | HS Chem. | Algebra I | none | Film/TV Production | Native |
| 10 | Intro Bio | Phys. II | Gen. Chem. II | Calculus I/College Stats | none | Biology | Native |
| 11 | HS Bio | HS Phys. | Gen. Chem. I | Algebra II/Pre-Calc. | 561 | Chemistry | ELL |
| 12 | Intro Bio | HS Phys. | Gen. Chem. I | Calculus I | none | Biology | ELL |

*calculus-based

*Data Analysis*

Student interviews were conducted via zoom and recorded using standard zoom features. Each interview was analyzed with the purpose of assigning specific levels of the cognitive framework to student responses following published procedure (Kaldaras & Wieman, 2023a). The first authors of the study used previously reported rubrics for Physics and Chemistry interview scenarios to assign levels to student responses. The inter-rater reliability of the rubrics was previously established (Kaldaras & Wieman, 2023a). Below we present the results of data analysis relevant for answering the RQs.

**Results**

*Using the Cognitive Framework to characterize student engagement in blended MSS*

Analysis of student interviews showed that students in this study exhibited similar response patterns to those in the prior study and these patterns related meaningfully to the sub-levels of the cognitive framework shown in Table 1. We were able to identify sample student responses at each sub-level of the framework as shown in Tables 6 and 7 for each disciplinary scenario. Since sample responses for each sub-level for Acceleration simulation were discussed in detail in Table 1, and sample responses in Table 6 parallel them for each sub-level, we will not discuss responses in Table 6 in greater detail, and focus on Table 7 sample responses for Chemistry context (Beer's law simulation).

*Table 6. Sample student responses at each sub-level of the framework for Acceleration.*

| | | |
|---|---|---|
| Conceptual | Mechanism | Acceleration is caused by net force applied over a certain mass: acceleration is directly related to net force reflected in the numerator, and inversely related to mass reflected in the denominator. |
| | Pattern | $F_{net}/m=a$ makes sense because acceleration has a direct linear relationship with net force and an inverse linear relationship with mass |
| | Description | It makes sense that $F_{applied}$ has to be subtracted from friction because friction takes away from the applied force and makes its effect on acceleration smaller |
| Quantitative | Mechanism | $F_{net}/m=a$ makes sense because every time I subtract applied force from friction and divide by the mass, that gives me the acceleration value |
| | Pattern | Adding 50 newtons to a given mass makes the acceleration increase by 1 m/s$^2$ |
| | Description | When applied force is 50 N acceleration is 3 m/s$^2$ |
| Qualitative | Mechanism | Applied force causes the object object of any mass to accelerate |
| | Pattern | Larger mass requires more applied force to get to the same acceleration |
| | Description | Applied force, friction and mass affect acceleration |

For Beer's Law simulation, level 1 (Qualitative) MSS is reflected in student ability to recognize that concentration, length of the container, type of substance and wavelength affect the absorbance ("Description" sub-level), describe qualitative patterns among the relevant variables such as the higher the concentration the higher the absorbance ("Pattern" sub-level), and recognizing that concentration is the main causal factor in changing the absorbance of a substance ("Mechanism" sub-level). Further, level 2 (Quantitative) MSS is reflected in students identifying specific numerical values of the relevant variables ("Description sub-level), describing specific quantitative patterns among the variables such as doubling the concentration and doubling the absorbance ("Pattern" sub-level) and finally developing a mathematical relationship for Beer's law grounded in specific values for the relevant variables ("Mechanism" sub-level). Notice that similarly to the examples for Acceleration, level 2 (quantitative) "Mechanism" is the lowest sub-level at which students start developing a mathematical relationship. Further, level 3 (Conceptual) "Description" is reflected in student ability to recognize the need to include unobservable variables, such as molar absorption coefficient (MAC) in order to fully account for the changes in absorbance for a given concentration, container width and wavelength and recognizing that MAC has to do with the properties of specific substances. Level 3 (Conceptual) "Pattern" is reflected in students ability to relate the quantitative patterns to the mathematical operations in the equation such as relating the direct linear relationship between absorbance on the one side and concentration and width of the container on the other side to the multiplication operation. Notice, this is in contrast to level 2 "Mechanism" reflecting the justification grounded in the logic of "numbers work out" type of response. Finally, at level 3 (Conceptual) "Mechanism" students describe the causal structure of the proposed equation by recognizing that the concentration is the main causal factor for changing absorbance and relating it to the equation structure by recognizing for example that when concentration is zero, absorbance is zero, which makes sense for multiplication operation.

*Table 7. Sample student responses for each sub-level of the framework for Beer's law.*

| **Conceptual** | Mechanism | Concentration causes absorbance because when concentration is zero absorbance is zero, which makes sense looking at the equation: multiplication by zero always gives zero. |
|---|---|---|
| | Pattern | $A = b \times c$ makes sense because doubling the concentration and the width of the container doubles the absorbance |
| | Description | MAC is important because it is unique to each substance and is a constant, and makes the equation work |
| **Quantitative** | Mechanism | A=e×b×c makes sense based on the data provided: if I multiply concentration by MAC it always gives me the absorbance because the length of the container is 1 |
| | Pattern | Doubling the concentration doubles the absorbancy |
| | Description | When concentration is 100 mM absorbance is 0.3 |
| **Qualitative** | Mechanism | Concentration causes changes in the absorbance of the substance |
| | Pattern | Larger concentrations have larger absorbancy |
| | Description | Concentration, container width and wavelength affect the absorbancy |

Notice that examples of student responses shown in Tables 6 and 7 reflect diverse ways students can engage in MSS at each sub-level. Student responses are grounded in their interactions with and observations of the corresponding PhET simulation. Even though we provided only one example per sub-level, there are a wide range of ways students can express their MSS at each sub-level that would be considered acceptable. When assigning a sub-level, we only care about the ideas expressed by students, not the form in which they are expressed. We don't expect academically correct standard English for any of the sub-levels. For example, students often explained that it is important to include friction into the equation because friction "takes away" from the applied force and makes its effect on acceleration smaller (see level 3 (Conceptual) "Description" sample response for Acceleration simulation). Even though it is not a standard scientific explanation, it is clear that the student understands both the scientific and the mathematical effect of friction on resulting acceleration. Therefore, this type of response would be aligned with the highest level of the framework. Further, many student examples shown in Tables 6 and 7 come from interviews with students whose native language is not English (see Table 5 for demographic information). Many of the phases shown in both tables are said with varying degrees of accent when speaking English. Further, you can see that students often use the word "absorbancy" instead of "absorbance", which is also acceptable even though "absorbancy" is not the correct scientific term. In short, student sample responses for the two interview scenarios shown in Tables 6 and 7 reflect a diverse range of ways students from various backgrounds can engage in blended MSS and the cognitive framework shown in Table 1 can be used to meaningfully characterize all these different forms of MSS. Specifically, we have found that the framework is as effective in characterizing student responses with this student population as it was in the previous study with students from dominant backgrounds. This answers RQ1 of the study.

*Student patterns of engagement in blended MSS*

We have found that the framework shown in Table 1 mostly functions as a learning progression (LP) with this student sample as well as with the previous one. Specifically, we have observed that all 12 students interviewed for this study tended to demonstrate engagement in blended MSS at all the levels below their highest demonstrated level. This was a consistent pattern with one notable exception as shown in Table 8, demonstration of level 2 (Quantitative) "Pattern" (sub-level 5 in Table 8). Table 8 shows the sub-levels (labeled 1-9) of the framework demonstrated by each study participant for the two disciplinary scenarios. Notice that most of the student responses are consistent with level 2 (Quantitative) and specifically sub-level "Pattern" (marked sub-level 5 in Table 8) of the cognitive framework. Further, as shown in Table 8 students demonstrate all types of blended MSS at level 2 and level 1 with the exception of level 2 (Quantitative) "Pattern" which only two demonstrate. This data suggests that while most of the students can develop the mathematical equation in both interview scenarios and justify this equation using specific numerical values of the relevant variables, they achieve that without engaging in quantitative pattern recognition consistent with level 2 (Quantitative) "Pattern". Note that all students were provided with the data along with the simulation (except student 11 who suggested the formulas based on the sims without needing to look at the data).

*Table 8. Patterns of engagement in blended MSS at each sub-level of the cognitive framework for Acceleration simulation (dark gray) and Beer's Law simulation (light gray) for all students. Note: white cells indicate levels not demonstrated by the student.*

| Level | Sub-level | Student |   |   |   |   |   |   |   |   |   |   |   |
|---|---|---|---|---|---|---|---|---|---|---|---|---|---|
|   |   | 1 | 2 | 3 | 4 | 5 | 6 | 7 | 8 | 9 | 10 | 11 | 12 |
| 3 | 9 |   |   |   |   |   |   |   | ■ |   |   | ▫ |   |
|   | 8 |   |   |   |   |   |   |   | ■▫ |   |   |   |   |
|   | 7 |   |   |   |   |   |   |   | ■▫ |   |   |   |   |
| 2 | 6 | ■▫ | ■▫ | ■▫ | ■▫ |   | ■▫ | ■▫ | ■▫ |   | ■▫ | ■▫ | ■▫ |
|   | 5 |   |   |   |   |   |   |   |   |   |   |   |   |
|   | 4 | ■▫ | ■▫ | ■▫ | ■▫ |   | ■▫ | ■▫ | ■▫ |   | ■▫ | ■▫ | ■▫ |
| 1 | 3 | ■▫ | ■▫ | ■▫ |   | ■▫ | ■▫ | ■▫ | ■▫ | ■▫ | ■▫ | ■▫ | ■▫ |
|   | 2 | ■▫ | ■▫ | ■▫ | ■▫ | ■▫ | ■▫ | ■▫ | ■▫ | ■▫ | ■▫ | ■▫ | ■▫ |
|   | 1 | ■▫ | ■▫ | ■▫ | ■▫ | ■▫ | ■▫ | ■▫ | ■▫ | ■▫ | ■▫ | ■▫ | ■▫ |

Therefore, the cognitive framework shown in Table 1 is functioning as an LP with the exception of this lack of evidence for engagement in level 2 "Pattern" type of MSS. Notice that this is only the case for students whose highest assigned level was level 2. Students who

14demonstrated MSS consistent with level 3 (student 8 for both scenarios and student 11 for Chemistry scenario) did demonstrate level 2 (Quantitative) "Pattern" MSS. Table 8 answers RQ2 of the study: students from diverse backgrounds interviewed in this study exhibit similar patterns of engagement in blended MSS as in the previous work with the one exception of lack of engagement in level 2 (Quantitative) "Pattern" for those students whose highest assigned level was level 2.

Students also tend to be consistent on sub-level assignments across the two disciplinary scenarios, which is consistent with the findings in the previous study (Kaldaras & Wieman, 2023a). This finding provides additional evidence towards our previous conclusion that blended MSS is a distinct cognitive construct likely unrelated to specific disciplinary context (Kaldaras & Wieman, 2023a). The only exception is student 11 who scored three sub-levels higher on the Chemistry scenario than on the Physics scenario. This student was completely unfamiliar with the Acceleration phenomenon but was a Chemistry major, and the Beer's Law phenomenon was easier for them to understand. This suggests that students might struggle to engage in higher level MSS in the contexts when the phenomenon is unfamiliar and challenging to understand even based on the sim, which is consistent with our prior findings (Kaldaras & Wieman, 2023a).

*Student Use of PhET Simulations*

We observed three different approaches in which PhET simulations helped facilitate MSS for students in this study. All three approaches reflect the degree to which PhET sims were helpful to students at varying levels of the framework in arriving at the correct math relationship and justifying it. Two of the approaches reflect a more frequent and deliberate use of PhET simulations by students in this sample compared to students from dominant backgrounds in the previous study. These two approaches were exhibited by students at level 2 of the framework.

One approach was similar for both student samples, and was exhibited by students at the highest level of the framework-level 3.

The first approach was common for students who were assigned level 2 "Mechanism" sub-level and proposed the formula based on looking at the data and interacting with the sim. These students did not need to see the list of possible formulas as shown in Table 9. The notable feature of this approach was that all these students (students 1-3, 5-7, 10-12 in Table 9) went back to the simulation to confirm their proposed mathematical relationship. When interacting with the simulation, they would focus on investigating how numerical values of the relevant variables change with respect to each other at various conditions and see if those changes are accurately described by their proposed math relationship. Students from dominant backgrounds in the prior study did not do that: they would stop at justifying their proposed math relationship using the data and did not go back to the sim to further confirm their formula. This demonstrates that level 2 students in this study use PhET simulations more often when making sense of the proposed formula and relating it back to the phenomenon, rather than relying on a table of data, compared to students from dominant backgrounds.

The second approach was similar to that observed for the majority of students in the previous study. The notable feature of this approach was that students at high levels of proficiency in MSS (such as level 3) were able to use the simulation alone to develop the mathematical relationship. In this study this approach was demonstrated by student 8 who was assigned the highest sub-level of the framework for both scenarios. This student did not need to see the list of formulas or the data to suggest mathematical relationships for Chemistry and Physics. Rather, this student used the simulation to develop both formulas. The student was able to notice the quantitative patterns among the relevant variables by interacting with the simulation, and proposed the formula based on the observed patterns. This finding suggests that for students with high proficiency in MSS (usually, level 3) PhET sims alone provide enough support to help them develop a mathematical relationship for the phenomenon. This trend seems to be holding for students from both dominant and underrepresented backgrounds. However, since in this study we only had one student to score in level 3, it would be beneficial to identify and interview additional students at this level to see whether this trend holds for a larger student population.

  The third approach was common for students who struggled to develop a mathematical relationship even when provided all the resources: the simulation, the data and the list of possible formulas. In the prior study these students would normally not be able to propose a possible mathematical relationship and their highest attained level would typically be level 1. In this study, however, some students who needed to see the list of possible formulas (such as students 7, 10 and 11 in Table 9) went back to the simulation and successfully narrowed down the list to the most likely formula. They were also able to use the simulation to justify their formula choice. For example, student 11, who demonstrated different highest levels of MSS on the two scenarios, needed to see the list of formulas for the Physics scenario, but not for Chemistry. Recall that student 11 also demonstrated less sophisticated MSS in Physics scenario. Importantly, this student was struggling to develop the mathematical relationship (F=ma or a/F/m) when interacting with the Acceleration simulation and seeing the data did not help figure out the relationship. However, when this student saw a list of possible formulas, they went back to the simulation to see if it could help them narrow down the correct formula. Using the simulation, the student quickly narrowed down the correct formula to two options: a=m/F and a = F/m. The student believed that the formula should have the form of a=...because the simulation is about what affects acceleration. This student indicated that division is the most likely operation but could not really explain why. The student continued to interact with the simulation and noticed that both friction and applied force affect acceleration and indicated that F in the list of possible formulas is likely the sum of forces, or F(applied)-F(friction). The student then showed with the example from the data how a=F/m is the correct formula by plugging in the numbers, which is consistent with level 2 "Mechanism". In short, the simulation seemed to be the most useful tool in helping this student make sense of the list of possible formulas, the data, and the phenomenon overall. This same exploration pattern was observed for two other level 2 students including students 7 and 10. Similar to student 11, both of these students were able to make sense of the list of possible formulas using the simulation, suggest the correct mathematical relationship and justify it using specific numerical values of the relevant variables. Without the simulation, it is unlikely that these students would engage in level 2 "Mechanism" MSS and develop the correct math relationship.

In contrast, students from dominant backgrounds in the previous study did not demonstrate this exploration pattern. Most students in the previous study who needed the formula list either were not able to narrow it down or used the list of possible formulas and went back to the provided data and plugged in the numbers to the formulas from the list. Since there were multiple formulas on the list that would yield the same numerical result, none of the students in the previous sample were able to accurately predict the exact math relationship based on the list of possible formulas. The student from the previous study never went back to the simulation to use it for narrowing down the list of possible formulas. This contrasting finding in the way students from the two studies used the simulations suggest that students in this study tend to use the simulation more. Furthermore, their exploration is more purposeful and productive since they seem to be using it specifically for making sense of their observations when narrowing down the list of possible formulas and succeeding in their exploration. However, notice that this is only the case for students at level 2 of the cognitive framework. Both students, whose highest demonstrated level was level 1 (students 4 and 9 in Table 6) needed to see the formula list, but were not able to use it or the simulation productively to narrow down the mathematical relationship. However, these students also went back to the simulation to try and narrow down the formula list, but they were not as successful as students who demonstrate level 2 MSS. Therefore, even though level 1 students from this study also tend to use the simulation more than those in the prior study, at this level they are not able to use it productively.

*Table 9. Students that needed to see the formula list and their sub-level assignment.*

| Student | Physics sub-level | Chemistry sub-level | Formulas for Physics | Formulas for Chemistry |
|---------|-------------------|---------------------|----------------------|------------------------|
| 1  | 2 (Mechanism) | 2 (Mechanism) | No  | No  |
| 2  | 2 (Mechanism) | 2 (Mechanism) | No  | No  |
| 3  | 2 (Mechanism) | 2 (Mechanism) | No  | No  |
| 4  | 1 (Pattern)   | 1 (Pattern)   | Yes | Yes |
| 5  | 2 (Mechanism) | 2 (Mechanism) | No  | No  |
| 6  | 2 (Mechanism) | 2 (Mechanism) | No  | No  |
| 7  | 2 (Mechanism) | 2 (Mechanism) | Yes | No  |
| 8  | 3 (Mechanism) | 3 (Mechanism) | No  | No  |
| 9  | 1 (Mechanism) | 1 (Mechanism) | Yes | Yes |
| 10 | 2 (Mechanism) | 2 (Mechanism) | Yes | Yes |
| 11 | 2 (Mechanism) | 3 (Mechanism) | Yes | No  |
| 12 | 2 (Mechanism) | 2 (Mechanism) | No  | No  |

Finally, it is important to point out that most students remembered talking about Acceleration or Beer's Law (although less often) at some point in their studies, but they did not recall any specifics, and never brought up Newton's Law ideas during the interview. Therefore, it was clear that the simulations supported autonomous exploration of the phenomenon with students having essentially no prior knowledge of the relevant content. These results help us answer RQ3 of the study: PhET simulations seem to not only help students from diverse backgrounds meaningfully engage in blended MSS across STEM disciplines with essentially no prior knowledge, but they seem to be more helpful to level 2 students in this sample compared to students from dominant backgrounds studies in the prior study (Kaldaras & Wieman, 2023a).

*Relationship between level placement and prior Math preparation*

The finding that level 2 students in this sample struggle to engage in level 2 (Quantitative) "Pattern" MSS made us wonder whether this pattern can potentially be correlated with prior Math preparation among these students. This was logical to suggest since engagement in level 2 "Pattern" MSS involves noticing specific quantitative patterns among the relevant variables and noticing linear, inverse and other relationships. Table 10 shows student highest level placement for both disciplinary scenarios, their Math coursework and Math SAT score.

*Table 10. Student highest level placement and their prior Math preparation.*

| Student | Physics level | Chemistry level | Highest Math Taken | Math SAT |
|---|---|---|---|---|
| 1 | 2 (Mechanism) | 2 (Mechanism) | College Stats. | ~500 |
| 2 | 2 (Mechanism) | 2 (Mechanism) | Differential equation and linear algebra | none |
| 3 | 2 (Mechanism) | 2 (Mechanism) | Calculus III | none |
| 4 | 1 (Pattern) | 1 (Pattern) | Calculus I | ~400 |
| 5 | 2 (Mechanism) | 2 (Mechanism) | Calculus I/College Stats. | 490 |
| 6 | 2 (Mechanism) | 2 (Mechanism) | Pre-Calc. | 500-600 |
| 7 | 2 (Mechanism) | 2 (Mechanism) | Calculus I/College Stats. | none |
| 8 | 3 (Mechanism) | 3 (Mechanism) | Higher than College Stats. | 680 |
| 9 | 1 (Mechanism) | 1 (Mechanism) | Algebra I | none |
| 10 | 2 (Mechanism) | 2 (Mechanism) | Calculus I/College Stats | none |
| 11 | 2 (Mechanism) | 3 (Mechanism) | Algebra II/Pre-Calc. | 561 |
| 12 | 2 (Mechanism) | 2 (Mechanism) | Calculus I | none |

Although the sample is small, Table 10 suggests that Math SAT in the ~400 range is the potential cut-off for level 1-level 2 transition. As you can see, the lowest Math SAT score for level 2 seems to be in the upper 400 range (student 5). It is harder to judge where the transition between level 2-level 3 occurs. This is consistent with our findings in the previous study: for level 2 and level 3 students, the Math SAT scores were above ~upper 400 but showed no correlation with level beyond that. This suggests that Math and Science dimensions are closely integrated at levels 2 and 3 of the framework and students need support in practicing to integrate the two dimensions to transition to higher levels. At level 1, however, it is likely that simply proficiency in Math is what prevents students from engaging in level 2 MSS. Further, data in Table 10 suggests that Algebra I is likely a cutoff course for level 1-level 2 transition. Notice that the highest Math course taken by level 1 students was Algebra I (student 9) and Calculus 1 (student 4). Logically, since the mathematical concepts evaluated in this study (such as linear and inverse relationships) relate to basic algebra, it is likely that Algebra I is a cutoff course for level 1-2 transition. As with Math SAT score, it is harder to identify the course related to level 2-3 transition. This is likely also due to the same reason- close integration of Math and Science dimensions at levels 2 and 3 of the framework.

**Discussion**

In this study we have shown that the cognitive framework for blended MSS previously validated with students from dominant backgrounds is applicable to characterizing engagement in MSS among undergraduate students from diverse backgrounds and institutions, including community colleges and universities serving predominantly underrepresented groups. This result is important in several ways. First, this finding suggests that the cognitive framework shown in Table 1 can be used to characterize engagement in MSS among students from highly diverse academic and cultural backgrounds. This is a valuable characteristic because this framework can potentially serve as an useful guide for creating effective learning opportunities for diverse students aligned to their individual learning needs. For example, we have used the framework to identify that lack of engagement in level 2 (Quantitative) "Pattern" type of MSS is a commonality among students from diverse backgrounds. It is likely that this struggle to engage in quantitative pattern identification is what prevents them from transitioning to level 3 type of blended MSS. This is consistent with our finding from a prior study showing that engagement in level 2 (Quantitative) "Pattern" is a necessary prerequisite for engaging in level 3 MSS (Kaldaras & Wieman, 2023a). Further, in a different study we have demonstrated that designing learning opportunities supporting the engagement in level 2 (Quantitative) "Pattern" MSS helps most undergraduate students from dominant backgrounds transition to level 3 of the cognitive framework (Kaldaras & Wieman, 2023b). Therefore, creating opportunities for diverse students focused on helping support their engagement in quantitative pattern identification is a promising strategy for helping these students transition to level 3. This strategy pairs PhET simulations with carefully scaffolded MC questions focused on guiding student exploration of the simulation to identify specific quantitative patterns among the relevant variables. Our previous results have shown that this approach helps support students in making sense of the quantitative patterns, which in turn helps them develop an accurate mathematical relationship focused on quantitative pattern identification (Kaldaras & Wieman, 2023b). We expect this would be a useful strategy with diverse students too.

Furthermore, we have seen in this study that diverse students at all levels of the framework tend to engage more with the simulation compared to students from dominant backgrounds. Based on student interaction with the simulations, it was clear that the simulations represented a straightforward and effective learning tool for engaging in MSS around the phenomenon. Data and lists of possible formulas were not as helpful to many without the simulation. The simulations seemed to be supporting effective autonomous student exploration of unfamiliar scientific phenomena. Therefore, when creating individualized learning opportunities for these students it is important to leverage this finding and scaffold student exploration of the simulation at all levels of the framework, but especially at level 2, since this seems to be the most critical level for helping transition to level 3.

The cognitive framework allows educators to easily build on the blended MSS demonstrated by students and guide the development of feasible learning opportunities. Specifically, it would be important to leverage the level 2 (Quantitative) "Description" type of MSS already demonstrated by all level 2 students in this sample to further guide them in identifying specific quantitative patterns among the numerical values of the relevant variables they have already identified to reach level 3.

Further, in terms of understanding how the proficiency in MSS develops, the finding that students can skip level 2 (Quantitative) "Pattern" but still demonstrate level 2 (Quantitative) "Mechanism" type of MSS suggests that students can effectively leverage level 2 (Quantitative) "Description" type of MSS to attaining level 2 (Quantitative) "Mechanism" sub-level. Specifically, they could leverage the numerical values of the relevant variables identified when engaging in level 2 "Description" to suggest a mathematical relationship among these variables based on numerical values alone, without identifying quantitative patterns. However, without engaging in level 2 "Pattern", students can't transition to level 3 of the framework. This an important finding because it highlights the importance of supporting students in learning to identify quantitative patterns to be able to transition to level 3 of the framework.

Finally, our data suggests that prior Math preparation tends to have more effect on student engagement in MSS at level 1 in terms of preventing students at level 1 from engaging at higher levels, and a lesser effect at levels 2 and 3. In terms of instruction to support students, these findings indicate that level 1 students need extensive support in both Math and the integration of Math with Science components to help them transition to level 2. Level 2 students for the most part need support in learning to uncover quantitative patterns, such as linear and inverse relationships and translate these relationships to specific mathematical operations in the formula to help them transition to level 3.

To conclude, the current study demonstrated the applicability of previously validated cognitive framework for MSS in characterizing MSS among diverse student populations and its potential in helping guide the development of individualized equitable learning opportunities aimed at helping these students develop higher proficiency in MSS with the help of PhET simulations.